\documentclass[oneside,10pt]{article}

\usepackage{times,url,mathrsfs}
\usepackage{amsmath}
\usepackage{amsfonts}
\usepackage{graphicx}
\usepackage{subfigure}
\newtheorem{lemma}{Lemma}

\begin{document}

\title{The Optimal Quantile Estimator for Compressed Counting}

\author{ Ping Li \hspace{0.01in} (pingli@cornell.edu) \hspace{0.2in} \\
       Cornell University,  Ithaca, NY 14853
  }
\date{}
\maketitle

\begin{abstract}
Compressed Counting (CC)\footnote{The results  were initially drafted in Jan 2008, as part of a report for private communications with several theorists. That report was later filed to arXiv\cite{Article:Li_CC_v0}, which, for shortening the presentation, excluded the content of the optimal quantile estimator. }
was recently proposed for very efficiently computing the (approximate) $\alpha$th frequency moments of data streams, where $0<\alpha \leq 2$. Several estimators were reported including the {\em geometric mean} estimator, the {\em harmonic mean} estimator, the {\em optimal power} estimator, etc. The {\em geometric mean} estimator is particularly interesting for theoretical purposes. For example, when $\alpha\rightarrow 1$, the complexity of CC (using the {\em geometric mean} estimator) is $O\left(1/\epsilon\right)$, breaking the well-known large-deviation bound $O\left(1/\epsilon^2\right)$. The case $\alpha\approx 1$ has  important applications, for example, computing entropy of data streams.

For practical purposes, this study proposes the {\em optimal quantile} estimator. Compared with previous estimators, this estimator is computationally more efficient and is also more accurate when $\alpha> 1$.

\end{abstract}

\section{Introduction}

\textbf{\em Compressed Counting (CC)}\cite{Article:Li_CC,Article:Li_CC_v0} was very recently proposed for efficiently computing the $\alpha$th frequency moments, where $0<\alpha\leq2$, in data streams. The underlying technique of CC is {\em maximally skewed stable random projections}, which significantly improves the well-know algorithm based on {\em symmetric stable random projections}\cite{Article:Indyk_JACM06,Proc:Li_SODA08}, especially when $\alpha\rightarrow 1$. CC boils down to a statistical estimation problem and various estimators have been  proposed\cite{Article:Li_CC,Article:Li_CC_v0}. In this study, we present an estimator based on the {\em optimal quantiles}, which is computationally more efficient and significantly more accurate when $\alpha> 1$, as long as the sample size is not too small.

One direct application of CC is to estimate entropy of data streams. A recent trend is to approximate entropy using frequency moments and estimate frequency moments using {\em symmetric stable random projections}\cite{Proc:Zhao_IMC07,Proc:Harvey_FOCS08}.  \cite{Report:Li_CC_entropy} applied CC to estimate entropy and demonstrated huge improvement (e.g., 50-fold) over previous studies.\\

CC was recently presented at {\em  MMDS 2008: Workshop on Algorithms for Modern Massive Data Sets}. Slides are available at \url{http://www.stanford.edu/group/mmds/slides2008/li.pdf}.

\subsection{The Relaxed Strict Turnstile Data Stream Model }

Compressed Counting (CC) assumes a {\em relaxed strict Turnstile} data stream model. In the {\em Turnstile} model\cite{Article:Muthukrishnan_05}, the input  stream $a_t = (i_t, I_t)$, $i_t\in [1,\ D]$ arriving sequentially describes the underlying signal $A$, meaning
\begin{align}
A_t[i_t] = A_{t-1}[i_t] + I_t,
\end{align} where the increment $I_t$ can be either positive (insertion) or negative (deletion). Restricting  $A_t[i]\geq 0$ at all  $t$  results in the {\em strict Turnstile} model, which suffices for describing most natural phenomena. CC constrains $A_t[i]\geq 0$ only at the $t$ we care about; however, when at $s \neq t$, CC allows $A_s[i]$ to be arbitrary.

Under the {\em relaxed strict Turnstile} model, the $\alpha$th frequency moment of a data stream $A_t$ is defined as
\begin{align}
F_{(\alpha)} = \sum_{i=1}^DA_t[i]^\alpha.
 \end{align}
When $\alpha = 1$, it is obvious that one can compute $F_{(1)} = \sum_{i=1}^D A_t[i] = \sum_{s=1}^t I_s$ trivially, using a simple counter.  When $\alpha \neq 1$, however, computing $F_{(\alpha)}$ exactly requires $D$ counters.

\subsection{Maximally-skewed Stable Random Projections}

Based on {\em maximally skewed stable random projections}), CC provides an very efficient mechanism for approximating $F_{(\alpha)}$.  One first generates a random matrix $\mathbf{R}\in\mathbb{R}^{D}$, whose entries are i.i.d. samples of a $\beta$-skewed $\alpha$-stable distribution with scale parameter 1, denoted by $r_{ij}\sim S(\alpha, \beta, 1)$.

By property of stable distributions\cite{Book:Zolotarev_86,Book:Samorodnitsky_94}, entries of the resultant projected vector $X = \mathbf{R}^\text{T}A_t\in\mathbb{R}^{k}$ are i.i.d. samples of a $\beta$-skewed $\alpha$-stable distribution whose scale parameter is the $\alpha$ frequency moment of $A_t$ we are after:
\begin{align}\notag
x_j = \left[\mathbf{R}^\text{T} A_t\right]_j = \sum_{i=1}^D r_{ij} A_t[i] \sim S\left(\alpha, \beta, F_{(\alpha)} = \sum_{i=1}^D A_t[i]^\alpha\right).
\end{align}

The skewness parameter $\beta \in[-1,1]$. CC recommends  $\beta = 1$, i.e., maximally-skewed, for the best performance.

In real implementation, the linear projection $X = \mathbf{R}^\text{T} A_t$ is conducted {\em incrementally}, using the fact that the {\em Turnstile} model is also linear. That is, for every incoming $a_t = (i_t, I_t)$, we update $x_j \leftarrow x_j + r_{i_tj} I_t$ for $j = 1$ to $k$.   This procedure is similar to that of {\em symmetric stable random projections}\cite{Article:Indyk_JACM06,Proc:Li_SODA08}; the difference is the distribution of the elements in $\mathbf{R}$.

\section{The Statistical Estimation Problem and Previous Estimators}

CC boils down to a statistical estimation problem. Given $k$ i.i.d. samples, $x_j \sim S\left(\alpha, \beta =1, F_{(\alpha)}\right)$, estimate the scale parameter $F_{(\alpha)}$.

\begin{figure}[h]
\begin{center}
\includegraphics[width = 3.5 in]{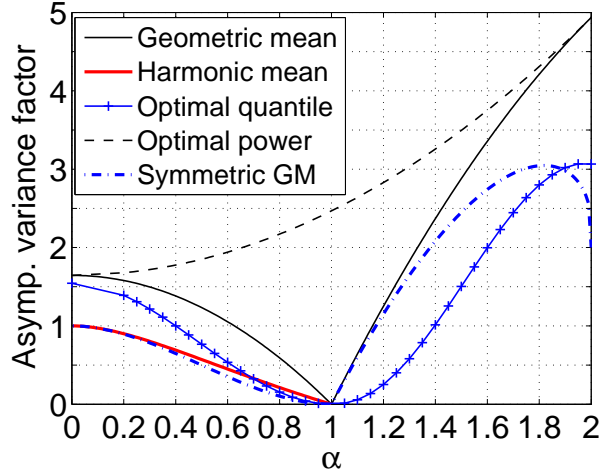}
\end{center}
\vspace{-0.3in}
\caption{Let $\hat{F}$ be an estimator of $F$ with asymptotic variance $\text{Var}\left(\hat{F}\right) = V\frac{F^2}{k} + O\left(\frac{1}{k^2}\right)$. We plot the $V$ values for the {\em geometric mean} estimator,  the {\em harmonic mean} estimator (for $\alpha<1$), the {\em optimal power} estimator (the lower dashed curve), and the {\em optimal quantile} estimator, along with the $V$ values for the {\em geometric mean} estimator for {\em symmetric stable random projections} in \cite{Proc:Li_SODA08} (``symmetric GM'', the upper dashed curve). When $\alpha\rightarrow 1$, CC achieves an ``infinite improvement'' in terms of the asymptotic variances.
}\label{fig_comp_var_factor}
\end{figure}

Assume $k$ i.i.d. samples $x_j \sim S\left(\alpha, \beta=1, F_{(\alpha)}\right)$. Various estimators were proposed in \cite{Article:Li_CC,Article:Li_CC_v0}, including the {\em geometric mean} estimator, the {\em harmonic mean} estimator, the {\em maximum likelihood} estimator, the {\em optimal quantile} estimator. Figure \ref{fig_comp_var_factor} compares their asymptotic variances along with the asymptotic variance of the {\em geometric mean} estimator for {\em symmetric stable random projections}\cite{Proc:Li_SODA08}.

\subsection{The geometric mean estimator, $\hat{F}_{(\alpha),gm}$, for $0<\alpha\leq 2$,  ($\alpha\neq 1$)}
\begin{align}\notag
&\hat{F}_{(\alpha),gm} = \frac{\prod_{j=1}^k |x_j|^{\alpha/k}} { \left(\cos^k\left(\frac{\kappa(\alpha)\pi}{2k}\right)/\cos \left(\frac{\kappa(\alpha)\pi}{2}\right)\right)  \left[\frac{2}{\pi}\sin\left(\frac{\pi\alpha}{2k}\right)
\Gamma\left(1-\frac{1}{k}\right)\Gamma\left(\frac{\alpha}{k}\right)\right]^k}.\\\notag
&\text{Var}\left(\hat{F}_{(\alpha),gm}\right) = \frac{F_{(\alpha)}^2}{k}\frac{\pi^2}{12}\left(\alpha^2+2-3\kappa^2(\alpha)\right)+O\left(\frac{1}{k^2}\right),\\\notag
&\kappa(\alpha) = \alpha, \ \ \ \text{if} \ \  \alpha <1, \hspace{0.2in} \kappa(\alpha) = 2-\alpha, \ \ \ \text{if} \ \  \alpha >1.
\end{align}
$\hat{F}_{(\alpha),gm}$ is unbiased and has exponential tail bounds for all $0<\alpha\leq 2$.

\subsubsection{The harmonic estimator, $\hat{F}_{(\alpha),hm,c}$, for $0<\alpha<1$}
\begin{align}\notag
&\hat{F}_{(\alpha),hm,c} = \frac{k\frac{\cos\left(\frac{\alpha\pi}{2}\right)}{\Gamma(1+\alpha)}}{\sum_{j=1}^k|x_j|^{-\alpha}}
\left(1- \frac{1}{k}\left(\frac{2\Gamma^2(1+\alpha)}{\Gamma(1+2\alpha)}-1\right) \right), \\\notag
&\text{E}\left(\hat{F}_{(\alpha),hm,c}\right) = F_{(\alpha)}+O\left(\frac{1}{k^2}\right),\hspace{0.5in}
\text{Var}\left(\hat{F}_{(\alpha),hm,c}\right) = \frac{F^{2}_{(\alpha)}}{k}\left(\frac{2\Gamma^2(1+\alpha)}{\Gamma(1+2\alpha)}-1\right) + O\left(\frac{1}{k^2}\right).
\end{align}
$\hat{F}_{(\alpha),hm,c}$ has exponential tail bounds.

\subsection{The maximum likelihood estimator, $\hat{F}_{(0.5),mle,c}$, for $\alpha = 0.5$ only}
\begin{align}\notag
&\hat{F}_{(0.5),mle,c}  =  \left(1-\frac{3}{4}\frac{1}{k}\right)\sqrt{\frac{k}{\sum_{j=1}^k\frac{1}{x_j}}},\\\notag
&\text{E}\left(\hat{F}_{(0.5),mle,c}\right) = F_{(0.5)} + O\left(\frac{1}{k^2}\right),\hspace{0.5in}
\text{Var}\left(\hat{F}_{(0.5),mle,c}\right) = \frac{1}{2}\frac{F_{(0.5)}^2}{k} + \frac{9}{8}\frac{F_{(0.5)}^2}{k^2} +
O\left(\frac{1}{k^3}\right).
\end{align}
$\hat{F}_{(0.5),mle,c}$ has exponential tail bounds.

\subsection{The optimal power estimator, $\hat{F}_{(\alpha),op,c}$, for $0<\alpha\leq 2$, ($\alpha\neq1$)} 
\begin{align}\notag
&\hat{F}_{(\alpha),op,c}
= \left(\frac{1}{k}\frac{\sum_{j=1}^k|x_j|^{\lambda^*
    \alpha}}{\frac{\cos\left(\kappa(\alpha)\frac{\lambda^*\pi}{2}\right) }{\cos^{\lambda^*}\left(\frac{\kappa(\alpha)\pi}{2}\right)
    }\frac{2}{\pi}
  \Gamma(1-\lambda^*)\Gamma(\lambda^*\alpha)\sin\left(\frac{\pi}{2}\lambda^*\alpha\right)} \right)^{1/\lambda^*}\times\\\notag
  &\hspace{0.5in} \left(1-\frac{1}{k}\frac{1}{2\lambda^*}\left(\frac{1}{\lambda^*}-1\right)\left(\frac{\cos\left(\kappa(\alpha){\lambda^*\pi}\right)\frac{2}{\pi}
  \Gamma(1-2\lambda^*)\Gamma(2\lambda^*\alpha)\sin\left(\pi\lambda^*\alpha\right)}{\left[\cos\left(\kappa(\alpha)\frac{\lambda^*\pi}{2}\right)\frac{2}{\pi}
  \Gamma(1-\lambda^*)\Gamma(\lambda^*\alpha)\sin\left(\frac{\pi}{2}\lambda^*\alpha\right)\right]^2}-1  \right) \right),\\\notag
&\text{E}\left(\hat{F}_{(\alpha),op,c}\right) = F_{(\alpha)} + O\left(\frac{1}{k^2}\right)\\\notag
&\text{Var}\left(\hat{F}_{(\alpha),op, c}\right)
=  F_{(\alpha)}^{2}\frac{1}{\lambda^*{^2}k}\left(\frac{\cos\left(\kappa(\alpha){\lambda^*\pi}\right)\frac{2}{\pi}
  \Gamma(1-2\lambda^*)\Gamma(2\lambda^*\alpha)\sin\left(\pi\lambda^*\alpha\right)}{\left[\cos\left(\kappa(\alpha)\frac{\lambda^*\pi}{2}\right)\frac{2}{\pi}
  \Gamma(1-\lambda^*)\Gamma(\lambda^*\alpha)\sin\left(\frac{\pi}{2}\lambda^*\alpha\right)\right]^2}-1  \right) + O\left(\frac{1}{k^2}\right).
\end{align}
\begin{align}\notag
\lambda^* = \underset{}{\text{argmin}}\ g\left(\lambda;\alpha\right), \hspace{0.2in}
g\left(\lambda;\alpha\right) =  \frac{1}{\lambda^2}\left(\frac{\cos\left(\kappa(\alpha){\lambda\pi}\right)\frac{2}{\pi}
  \Gamma(1-2\lambda)\Gamma(2\lambda\alpha)\sin\left(\pi\lambda\alpha\right)}{\left[\cos\left(\kappa(\alpha)\frac{\lambda\pi}{2}\right)\frac{2}{\pi}
  \Gamma(1-\lambda)\Gamma(\lambda\alpha)\sin\left(\frac{\pi}{2}\lambda\alpha\right)\right]^2}-1  \right).
\end{align}

When $0<\alpha<1$,  $\lambda^*<0$ and $\hat{F}_{(\alpha),op,c}$ has exponential tail bounds.

$\hat{F}_{(\alpha),op,c}$ becomes the {\em harmonic mean} estimator when $\alpha =0+$, the {\em arithmetic mean} estimator when $\alpha = 2$, and the {\em maximum likelihood} estimator when $\alpha = 0.5$.

\section{The Optimal Quantile Estimator}

Because $X \sim S\left(\alpha,\beta=1,F_{(\alpha)}\right)$ belongs to the location-scale family (location is zero always), one can estimate the scale parameter $F_{(\alpha)}$ simply from the sample qantiles.

\subsection{A General Quantile Estimator}
Assume $x_j \sim S\left(\alpha, 1, F_{(\alpha)}\right)$, $j = 1$ to $k$.  One possibility is to use the $q$-quantile of the absolute values, i.e.,
\begin{align}\label{eqn_quantile}
\hat{F}_{(\alpha),q} = \left(\frac{q\text{-Quantile}\{|x_j|, j = 1, 2, ..., k\}}{W_q}\right)^\alpha.
\end{align}
where
\begin{align}\label{eqn_W}
W_q = q\text{-Quantile}\{|S(\alpha,\beta=1,1)|\}.
\end{align}

Denote $Z = |X|$, where $X\sim S\left(\alpha,1,F_{(\alpha)}\right)$. Note that when $\alpha <1$, $Z=X$. Denote the probability density function of $Z$ by $f_Z\left(z;\alpha,F_{(\alpha)}\right)$, the probability cumulative function by $F_Z\left(z;\alpha,F_{(\alpha)}\right)$, and the inverse cumulative function by $F_Z^{-1}\left(q;\alpha,F_{(\alpha)}\right)$.

We can analyze the asymptotic (as $k\rightarrow \infty$) variance of $\hat{F}_{(\alpha),q}$, presented in Lemma \ref{lem_q_var}.
\begin{lemma}\label{lem_q_var}
\begin{align}\label{eqn_q_var}
\text{Var}\left(\hat{F}_{(\alpha),q}\right)
=&\frac{1}{k}\frac{(q-q^2)\alpha^2}{f^2_Z\left(F_Z^{-1}\left(q;\alpha,1\right);\alpha,1\right) \left(F_Z^{-1}\left(q;\alpha,1\right)\right)^2 } F_{(\alpha)}^2+O\left(\frac{1}{k^2}\right).
\end{align}
\textbf{Proof:} \ \ The proof directly follows from known statistical results on sample quantiles, e.g., \cite[Theorem
9.2]{Book:David}, and the ``delta'' method.
\begin{align}\notag
\text{Var}\left(\hat{F}_{(\alpha),q}\right)
=& \frac{1}{k}\frac{q-q^2}{f^2_Z\left(F_Z^{-1}\left(q;\alpha,F_{(\alpha)}\right);\alpha,F_{(\alpha)}\right) \left(F_Z^{-1}\left(q;\alpha,1\right)\right)^2 } \left(F_{(\alpha)}\right)^{\left((\alpha-1)/\alpha\right)^2}\alpha^2+O\left(\frac{1}{k^2}\right)\\\notag
=&\frac{1}{k}\frac{(q-q^2)\alpha^2}{f^2_Z\left(F_Z^{-1}\left(q;\alpha,1\right);\alpha,1\right) \left(F_Z^{-1}\left(q;\alpha,1\right)\right)^2 } F_{(\alpha)}^2+O\left(\frac{1}{k^2}\right),
\end{align}
\noindent using the fact that
\begin{align}\notag
F_Z^{-1}\left(q; \alpha, F_{(\alpha)}\right) = F_{(\alpha)}^{1/\alpha} F_Z^{-1}\left(q; \alpha,1\right), \hspace{0.5in}
f_Z\left(z; \alpha, F_{(\alpha)}\right) = F_{(\alpha)}^{-1/\alpha} f_Z\left(z\alpha^{-1/\alpha}; \alpha, 1\right).
\end{align}
\end{lemma}

We can choose $q=q^*$ to minimize the asymptotic variance factor,
$\frac{(q-q^2)\alpha^2}{f^2_Z\left(F_Z^{-1}\left(q;\alpha,1\right);\alpha,1\right) \left(F_Z^{-1}\left(q;\alpha,1\right)\right)^2 }$, which is apparently a convex function of $q$, although there appears no simple algebraic method to prove it (except when $\alpha = 0+$).

We denote the optimal quantile estimator as $\hat{F}_{(\alpha),oq} = \hat{F}_{(\alpha),q^*}$.

\subsection{The Optimal Quantiles}

The optimal quantiles, denoted by $q^* = q^*(\alpha)$, has to be determined by numerical procedures, using the simulated probability density functions for stable distributions. We used the \textbf{fBasics} package in \textbf{R}. We, however, found those functions had numerical problems when $1<\alpha<1.011$ and $0.989<\alpha<1$.

For all other estimators, we have not noticed any numerical issues even when $\alpha = 1-10^{-4}$ or $1+10^{-4}$. Therefore, we do not consider there is any numerical instability for CC, as far as the method itself is concerned.

Table \ref{tab_oq} presents the numerical results, including $q^*$,  $W_{q^*} = q^*\text{-Quantile}\{|S(\alpha,\beta=1,1)|\}$, and the variance of $\hat{F}_{(\alpha),oq}$ (without the $\frac{1}{k}$ term). The variance factor is also plotted in Figure \ref{fig_comp_var_factor}, indicating significant improvement over the geometric mean estimator when $\alpha>1$.

\begin{table}[h]
\caption{\small
 }
\begin{center}{\scriptsize
\begin{tabular}{l l l l}
\hline \hline
$\alpha$ &$q^*$  &Var  &$W_{q^*}$ \\\hline
0.20 &0.180& 1.39003806& 0.05561700\\
0.30 &0.167& 1.21559359& 0.11484008\\
0.40 &0.151& 1.00047427& 0.2720723\\
0.50 &0.137& 0.76653704& 0.4522449\\
0.60 &0.127& 0.53479789& 0.7406894\\
0.70 &0.116& 0.32478420& 1.231919\\
0.80 &0.108& 0.15465894& 2.256365\\
0.85 &0.104& 0.08982992& 3.296870\\
0.90 &0.101& 0.04116676& 5.400842\\
0.95 &0.098& 0.01059831 &1.174773\\
0.96 &0.097& 0.006821834 & 14.92508\\
0.97 &0.096& 0.003859153 &20.22440\\
0.98   &0.0944& 0.001724739&    30.82616\\
0.989 & 0.0941& 0.0005243589& 56.86694\\
1.011 & 0.8904& 0.0005554749& 58.83961\\
1.02 & 0.8799& 0.001901498&  32.76892\\
1.03  & 0.869& 0.004424189& 22.13097\\
1.04 &0.861& 0.008099329& 16.80970\\
1.05 & 0.855& 0.01298757& 13.61799\\
1.10 &0.827& 0.05717725& 7.206345\\
1.15 &0.810& 0.1365222& 5.070801\\
1.20 &0.799& 0.2516604& 4.011459\\
1.30 &0.784& 0.5808422& 2.962799\\
1.40 &0.779& 1.0133272& 2.468643\\
1.50 &0.778& 1.502868& 2.191925\\
1.60 &0.785 &1.997239& 2.048035\\
1.70 & 0.794 &2.444836& 1.968536\\
1.80 &0.806 &2.798748& 1.937256\\
1.90 &0.828 &3.019045& 1.976624\\
2.00 &0.862 &3.066164& 2.097626\\
\hline\hline
\end{tabular}
}
\end{center}
\label{tab_oq}
\end{table}

\subsection{Comments on the Optimal Quantile Estimator}

The optimal quantile estimator has at least two advantages:
\begin{itemize}
\item When the sample size $k$ is not too small (e.g., $k\geq 50$), $\hat{F}_{(\alpha),oq}$ is more accurate then $\hat{F}_{(\alpha),gm}$, especially for $\alpha>1$.
\item $\hat{F}_{(\alpha),oq}$ is computationally more efficient.
\end{itemize}

The disadvantages are:
\begin{itemize}
\item For small samples (e.g., $k\leq20$), $\hat{F}_{(\alpha),oq}$ exhibits bad behaviors when $\alpha>1$.
\item Its theoretical analysis, e.g., variances and tail bounds, is based on the density function of skewed stable distributions, which do not have closed-forms. The tail bound bounds can be obtained similarly using the method developed in \cite{Article:Li_arXiv_sym_oq}.
\item The important parameters, $q^*$ and $W_{q^*}$, are obtained from the numerically-computed density functions. Due to the numerical difficulty in those functions, we can only obtain $q^*$ and $W_{q^*}$ values for $\alpha\geq 1.011$ and $\alpha\leq 0.989$.
\end{itemize}

\section{Conclusion}

Compressed Counting (CC) dramatically improves {\em symmetric stable random projections}, especially when $\alpha\approx 1$, and has important applications in data streams computations such as entropy estimation.

CC boils down to a statistical estimation problem. We propose the optimal quantile estimator, which considerably improves the previously proposed geometric mean estimator when $\alpha>1$, at least asymptotically. For practical purposes, this estimator should be very useful. However, for theoretical purposes,  it can not replace the geometric mean estimator.

\appendix


{\small
\begin{thebibliography}{10}

\bibitem{Book:David}
Herbert~A. David.
\newblock {\em Order Statistics}.
\newblock John Wiley \& Sons, Inc., New York, NY, second edition, 1981.

\bibitem{Proc:Harvey_FOCS08}
Nicholas J.~A. Harvey, Jelani Nelson, and Krzysztof Onak.
\newblock Sketching and streaming entropy via approximation theory.
\newblock In {\em FOCS}, 2008.

\bibitem{Article:Indyk_JACM06}
Piotr Indyk.
\newblock Stable distributions, pseudorandom generators, embeddings, and data
  stream computation.
\newblock {\em Journal of ACM}, 53(3):307--323, 2006.

\bibitem{Article:Li_CC}
Ping Li.
\newblock Compressed counting.
\newblock {\em CoRR}, abs/0802.2305, 2008.

\bibitem{Article:Li_arXiv_sym_oq}
Ping Li.
\newblock Computationally efficient estimators for dimension reductions using
  stable random projections.
\newblock {\em CoRR}, abs/0806.4422, 2008.

\bibitem{Proc:Li_SODA08}
Ping Li.
\newblock Estimators and tail bounds for dimension reduction in $l_\alpha$
  ($0<\alpha\leq 2$) using stable random projections.
\newblock In {\em SODA}, pages 10 -- 19, 2008.

\bibitem{Article:Li_CC_v0}
Ping Li.
\newblock On approximating frequency moments of data streams with skewed
  projections.
\newblock {\em CoRR}, abs/0802.0802, 2008.

\bibitem{Report:Li_CC_entropy}
Ping Li.
\newblock A very efficient scheme for estimating entropy of data streams using
  compressed counting.
\newblock Technical report, Department of Statistical Science, Cornell
  University, 2008.

\bibitem{Article:Muthukrishnan_05}
S.~Muthukrishnan.
\newblock Data streams: Algorithms and applications.
\newblock {\em Foundations and Trends in Theoretical Computer Science},
  1:117--236, 2 2005.

\bibitem{Book:Samorodnitsky_94}
Gennady Samorodnitsky and Murad~S. Taqqu.
\newblock {\em Stable Non-Gaussian Random Processes}.
\newblock Chapman \& Hall, New York, 1994.

\bibitem{Proc:Zhao_IMC07}
Haiquan Zhao, Ashwin Lall, Mitsunori Ogihara, Oliver Spatscheck, Jia Wang, and
  Jun Xu.
\newblock A data streaming algorithm for estimating entropies of od flows.
\newblock In {\em IMC}, San Diego, CA, 2007.

\bibitem{Book:Zolotarev_86}
Vladimir~M. Zolotarev.
\newblock {\em One-dimensional Stable Distributions}.
\newblock American Mathematical Society, Providence, RI, 1986.

\end{thebibliography}

}
\end{document}